\documentclass[showpacs,preprintnumbers,showkeys,amsmath,amssymb]{revtex4}
\usepackage{graphicx}
\usepackage{dcolumn}
\usepackage{bm}
\begin{document}

\preprint{}

\title{Hybrid quantum gates between flying photon and diamond nitrogen-vacancy centers assisted by optical microcavities}

\author{Hai-Rui Wei \& Gui Lu Long\footnote{Correspondence author: gllong@tsinghua.edu.cn) } }

\address{State Key Laboratory of Low-Dimensional Quantum Physics and Department of Physics, Tsinghua University, Beijing 100084, China}

\begin{abstract}
Hybrid quantum gates hold great promise for quantum information processing since they preserve the advantages of different quantum systems. Here we present compact quantum circuits to deterministically implement controlled-NOT,  Toffoli, and Fredkin gates between a flying photon qubit and diamond nitrogen-vacancy (NV) centers assisted by  microcavities. The target qubits of these universal quantum gates are encoded on the spins of the electrons associated with the diamond NV centers and they have long coherence time for storing information, and the control qubit is encoded on the polarizations of the flying photon and can be easily manipulated. Our quantum circuits are compact, economic, and simple. Moreover, they do not require additional qubits. The complexity of our schemes for universal three-qubit  gates is much reduced, compared to  the synthesis  with two-qubit entangling gates.  These schemes have high fidelities and efficiencies, and they are feasible in experiment.
\end{abstract}



\maketitle

\section{Introduction}

A quantum computer\cite{book} is more powerful than a classical computer in solving certain computationally demanding tasks. Quantum logic gates are the  fundamental  building blocks of a quantum computer, and a quantum computing task can be completed using a sequence of quantum gates as described in a  quantum circuit. It is well known that any quantum computing can be decomposed into a sequence of single-qubit gates and two-qubit entangling gates\cite{uni}, and analytical expressions\cite{liuy} for an arbitrary $n$-qubit unitary gate have been explicitly derived using the methods provided in Ref. \cite{uni}.  One of the most popular universal quantum gates
is the controlled-NOT (CNOT) gate. Quantum circuit received great attention over the years, in particular the CNOT gate (or the controlled phase gate)\cite{uni,liuy,3CNOT2,Theor,Bull0,Bull,longprl,xug1,xug2} and the hyperparallel CNOT gate\cite{renCNOT,renCNOT2,RenCNOTpra}. The theoretical lower bound of an unstructured $n$-qubit quantum computation is $(4^n - 3n - 1)/4$  CNOT gates\cite{Theor}. In multi-qubit systems, the fundamental three-qubit Toffoli gate\cite{Toffoli} or Fredkin gate\cite{Fredkin} form a family of universal quantum gates with the help of Hadamard operations, and they are valuable in fault-tolerant quantum circuits and some quantum algorithms. The realization of a Toffoli gate or a Fredkin gate in terms of
two-qubit entangling gates is troublesome  as the optimal cost is six CNOT gates\cite{Toffolicost} for a Toffoli gate and five two-qubit entangling gates for a Fredkin gate\cite{Fredkincost}. It is desirable to seek efficient schemes for directly implementing the Toffoli and Fredkin gates so as to  speedup the quantum computation.



A single photon is a perfect information carrier  and it has a flexible controllability. However, it seems  unsuitable for quantum computing as  the direct interaction between individual photons is very weak. Different to photonic qubit\cite{Bull1,Bull3}, matter qubits, such as atoms, quantum dots (QDs), superconduction  junctions,  and diamond nitrogen-vacancy (NV) defect centers,  are widely utilized in quantum computing because of their long-lived coherence time and their good scalability. Compared with other candidates, a diamond NV center is a particularly promising one for a qubit as it has an ultralong coherence time (1.8 ms)\cite{coherence1} even at the room temperature. In a diamond NV center, the electron spin can be exactly populated by the optical pumping with 532 nm light\cite{population}, and it can be manipulated\cite{population,manipulate2,manipulate4,manipulate5} and
readout\cite{readout2,readout3} by using the microwave excitation.  The techniques to transfer the information from electron spins to  nuclear spins were developed well\cite{register1,register3,register4}.  Besides, some important tasks in quantum computation have been investigated and even been realized in experiment on diamond NV centers. For example, in 2004, Jelezko \emph{et al.}\cite{CROT} carried out the experiments for implementing the hybrid controlled-ROT gate on an electron-nuclear system. In 2012, Sar \emph{et al.}\cite{decoherence-protected} realized the decoherence-protected conditional rotation gates on hybrid electron-nuclear systems. In 2010, Yang \emph{et al.}\cite{CCPF} proposed a conditional phase gate on three diamond NV centers. In 2013, Wei and Deng\cite{our} proposed some compact schemes for implementing universal gates on diamond NV centers, and Wang \emph{et al.}\cite{Wangchuan} designed a quantum circuit for the photonic controlled phase gate via a diamond NV center. In 2015, Ren, Wang and Deng\cite{RenCNOTpra} presented
the dipole induced transparency  of a diamond NV center embedded in a
photonic crystal cavity coupled to two waveguides, and proposed two universal hyperparallel hybrid photonic quantum logic gates, including a hybrid hyper-controlled-NOT gate and a hybrid hyper-Toffoli gate, on
photon systems in both the polarization and the spatial-mode degrees of freedom, which can be used to perform
more quantum operations with less resources and depress the resources consumed and the photonic dissipation.  Recently, some interesting works for quantum information processing  have been achieved on diamond NV centers, such as entanglement generation\cite{photon-NV,nuclear-nuclear,NV-NV2,NV-NV3,NV-NV5,threemeters,entanglement-by-measurement}, quantum manipulation\cite{cpb1,apl1,light}, quantum teleportation between solid-state qubits separated by three meters\cite{teleportation}, and hyperentanglement\cite{ren} and entanglement\cite{wangchuanscpma,shengybcsb} purification and concentration.



A light-matter system\cite{Hu1,Hu2,Bonato,atom1} coupled to a cavity provides an important platform to study quantum information processing. For example, some important schemes for the conventional parallel quantum computation\cite{atom1,weioe,HuaMPRA,HuaSR} or the hyperparallel photonic quantum computation\cite{renCNOT,renCNOT2,RenCNOTpra}  were proposed with the light-matter platform coupled to optical microcavities. By using a flying photon as a bus, schemes for universal gates on atoms\cite{atom2} and QDs\cite{weioe20014} have been proposed. Hybrid quantum gates on two or more physical systems inherit all the advantages of the different systems.

In this paper, we focus on designing compact quantum circuits to implement CNOT, Toffoli, and Fredkin gates between a flying photon and  solid-state diamond NV centers coupled to cavities. These quantum circuits are constructed by utilizing the input-output process of the single photon as a result of cavity quantum electrodynamics and optical spin selection rules. The schemes well work at the degeneracy of the spin 1 system and the gate's mechanism is deterministic in principle.  The target qubits are encoded on the ground states of the electrons $|m_s=\pm1\rangle$  associated with the diamond NV centers. The control qubit is encoded on the polarizations of the flying single photon.  Our schemes have some advantages. First,  our quantum circuits for these universal quantum   gates are compact and economic. Second,  they do not require additional qubits.  Third, the control qubit is the flying photon which has the flexible controllability. Fourth, the target qubits are encoded on the spins of the electrons associated with NV centers which have the relatively long coherence time even at room temperature and are perfect for the storage of quantum information. Fifth, the complexity of our schemes for three-qubit quantum gates beats their synthesis procedures largely. The high fidelities and efficiencies of our schemes show that they  may be feasible with current technology.

\section{Results}\label{sec2}
{\bf A diamond nitrogen-vacancy center confined in an optical resonant microcavity.}
A diamond NV center consists of a vacancy adjacent to a substitutional nitrogen atom (typically $^{14}$N). In  a  diamond NV center, both the nuclear spins (typically $^{13}$C with $I$=1/2 or $^{14}$N with $I$=1) and the electron spins are promising for quantum information processing. The ground states of the electron, $|0\rangle\equiv|m_s=0\rangle$ and the two-fold degenerate states $|\pm\rangle\equiv|m_s=\pm1\rangle$, is split by $D\approx 2.87$ GHz in a zero external field due to the spin-spin interaction\cite{split}. The six excited states\cite{photon-NV}
$|A_1\rangle=(|E_-\rangle|+\rangle-|E_+\rangle|-\rangle)/\sqrt{2}$,
$|A_2\rangle=(|E_-\rangle|+\rangle+|E_+\rangle|-\rangle)/\sqrt{2}$,
$|E_x\rangle=|X\rangle|0\rangle$, $|E_y\rangle=|Y\rangle|0\rangle$,
$|E_1\rangle=(|E_-\rangle|-\rangle-|E_+\rangle|+\rangle)/\sqrt{2}$,
and
$|E_2\rangle=(|E_-\rangle|-\rangle+|E_+\rangle|+\rangle)/\sqrt{2}$
are dominated by the NV center's C$_{3v}$ symmetry and the spin-spin, spin-orbit interactions without external strain and electric or magnetic fields. Here $|E_\pm\rangle$, $|X\rangle$, and $|Y\rangle$ are the orbital states of an NV center. The spin-orbit interaction (5.5 GHz)\cite{split1,split2} splits the excited states into three two-fold degeneracy pairs ($A_1$, $A_2$) (to be shifted up), ($E_x$, $E_y$), and ($E_1$, $E_2$) (to be shifted down). The
spin-spin interaction (1.42 GHz) shifts up states ($A_1$, $A_2$, $E_1$, and $E_2$) by 1.42/3 GHz and shifts down states ($E_x$, $E_y$) by $2*1.42/3$ GHz\cite{split1,split2}. Besides, it splits $A_2$ and $A_1$ by $\pm$ 1.55 GHz\cite{split1,split2}. The local non-axial high strain (10 GHz, larger than the spin-orbit splitting in the presence of the zero field) splits the excited states into two branches, ($A_2$, $A_1$, and $E_x$) and ($E_y$, $E_1$, and $E_2$). The state $|A_2\rangle$ is robust against the relatively small strain and magnetic fields with the stable symmetry properties, preserving the polarization properties of its optical transitions. The frequency of the spin-selective optical resonant transition can be tuned via an application of a controlled external electric field\cite{split2,electricfield,electricfield1,electricfield2,electricfield3}. In 2011, Bassett \emph{et al.}\cite{electricfield3} experimentally demonstrated an exceeding 10 GHz optical transition frequency. The transitions between the ground states are in the microwave frequency regime, and the transitions between the ground states and the excited states are in the optical regime. With microwave and laser, one can prepare, store, and read out the states of the solid-state electron spins\cite{PRL}. Here we encode the qubit on the sublevels $|\pm\rangle$, and take $|A_2\rangle$ as an auxiliary state. $|A_2\rangle$ decays into $|\pm\rangle$ with the right-circularly-polarized ($R$) and left-circularly-polarized ($L$) photons [see Fig. 1b],  respectively, owning to total angular momentum conservation.  They take place with the equal probability.

\begin{figure}[!h]           
\begin{center}
\includegraphics[width=8.2 cm,angle=0]{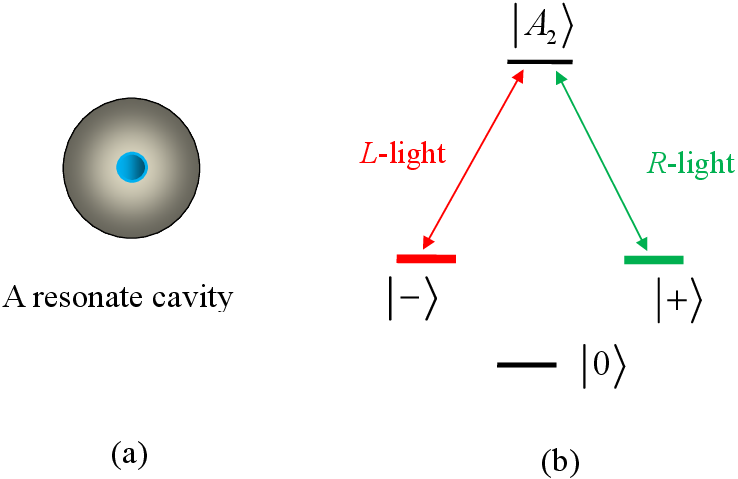}
\caption{ (a) Schematic diagram of an NV-cavity system. (b) The energy-level diagram of an NV-cavity system. The triple ground states $|\pm\rangle\equiv|m_s=\pm1\rangle$ are chosen to act as the two levels for an electron-spin qubit. The excited state $|A_2\rangle=(|E_-\rangle|+1\rangle+|E_+\rangle|-1\rangle)/\sqrt{2}$ is an auxiliary state. $|\pm\rangle\rightarrow|A_2\rangle$ are driven by the right- and left- circularly-polarized photons, respectively.
} \label{level}
\end{center}
\end{figure}

In 2011, Chen \emph{et al.}\cite{NV-NV3} discussed a composite unit, that is,  a diamond NV center confined inside a single-sided resonator [see Fig. 1a]. Combing the Heisenberg equations of motion\cite{QObook}
\begin{eqnarray}       \label{eq1}
\begin{split}
&\frac{d\hat{a}}{dt}  = -\left[i(\omega_c-\omega_p)+\frac{\kappa}{2}\right]\hat{a}(t)-\emph{\text{g}}\sigma_{-}(t) - \sqrt{\kappa}\hat{a}_{in}, \\
&\frac{d\sigma_-}{dt} =
-\left[i(\omega_{0}-\omega_p)+\frac{\gamma}{2}\right]\sigma_{-}(t)-\emph{\text{g}}\sigma_z(t)\hat{a}(t)
+ \sqrt{\gamma}\sigma_z(t)\hat{b}_{in}(t), 
\end{split}
\end{eqnarray}
and the standard input-output relation for the cavity
\begin{eqnarray}       \label{eq2}
\hat{a}_{out} = \hat{a}_{in}+ \sqrt{\kappa}\hat{a}(t),
\end{eqnarray}
 the explicit expression of the reflection coefficient for the NV-cavity unit in the weak excitation limit $\langle\sigma_z\rangle=-1$ can be written as\cite{NV-NV3,Hersenberg}
\begin{eqnarray}       \label{eq3}
r(\omega_p)=\frac{[i(\omega_{c}-\omega_p)-\frac{\kappa}{2}][i(\omega_{0}-\omega_p)+\frac{\gamma}{2}]+\emph{\text{g}}^2}
           {[i(\omega_{c}-\omega_p)+\frac{\kappa}{2}][i(\omega_{0}-\omega_p)+\frac{\gamma}{2}]+\emph{\text{g}}^2}.
\end{eqnarray}
Here $\hat{a}$ and $\sigma_-$ are the annihilation operator of the cavity mode and the transition operator of the diamond NV center with the frequencies $\omega_c$ and $\omega_0$, respectively. $\omega_p$ is the frequency of the input single photon. $\sigma_z(t)$ presents the inversion operator of the NV center. $\gamma$ and $\kappa$ are the NV decay rate and the cavity damping rate, respectively. $\emph{\text{g}}$ is the coupling strength between an NV center and a cavity. The vacuum input field $b_{in}(t)$ has the commutation relation $[\hat{b}_{in}(t),\hat{b}_{in}^\dag(t')]=\delta(t-t')$.

When the diamond NV center  confined inside a resonant  cavity interacts with a resonant  single photon, i.e., $\omega_0=\omega_c=\omega_p$,  the reflection coefficients for the hot cavity ($\emph{\text{g}}\neq0$) and the cold cavity ($\emph{\text{g}}=0$) can be written as
\begin{eqnarray}       \label{eq4}
r=\frac{-\kappa\gamma+4\emph{\text{g}}^2}{\kappa\gamma+4\emph{\text{g}}^2},\qquad\qquad\qquad
r_0=-1.
\end{eqnarray}
That is, the change of the incident photon can be summarized as follows\cite{NV-NV3}:
\begin{eqnarray}       \label{eq5}
\begin{split}
|R\rangle|+\rangle \;\xrightarrow{\text{NV}}\;|r||R\rangle|+\rangle,\;\;\;\;
|L\rangle|-\rangle\;\xrightarrow{\text{NV}}\; |r||L\rangle|-\rangle,\;\;\;
|R\rangle|-\rangle\;\xrightarrow{\text{NV}}\;-|R\rangle|-\rangle,\;\;\;\;
|L\rangle|+\rangle\;\xrightarrow{\text{NV}}\;-|L\rangle|+\rangle.
\end{split}
\end{eqnarray}
When $\emph{\text{g}}\geq5\sqrt{\gamma\kappa}$ and $\omega_0=\omega_c=\omega_p$, $r(\omega_p)\simeq1$ and $r_0(\omega_p) =-1$\cite{NV-NV3}, and Eq. (\ref{eq5}) becomes
\begin{eqnarray}       \label{eq6}
\begin{split}
|R\rangle|+\rangle \;\xrightarrow{\text{NV}}\;|R\rangle|+\rangle, \;\;\;\;
|L\rangle|-\rangle \;\xrightarrow{\text{NV}}\;|L\rangle|-\rangle,\;\;\;\;
|R\rangle|-\rangle\;\xrightarrow{\text{NV}}\;&-|R\rangle|-\rangle,\;\;\;
|L\rangle|+\rangle\;\xrightarrow{\text{NV}}\;-|L\rangle|+\rangle.
\end{split}
\end{eqnarray}

{\bf Compact quantum circuit for implementing a hybrid CNOT gate.}
The framework of our proposal for implementing a CNOT gate is shown in  Fig. 2. It performs a not operation on the diamond NV center when the flying single photon is in state $|L\rangle$. Let us describe its principle in detail as follows.

\begin{figure}[h!]      
\begin{center}
\includegraphics[width=8 cm,angle=0]{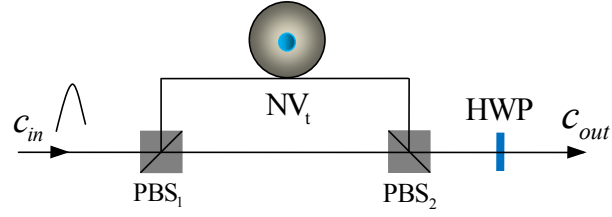}
\caption{ Compact quantum circuit for implementing the CNOT gate on a hybrid photon-NV system with the flying single photon polarization as the control qubit and the electron spin in the  diamond NV center as the target qubit. PBS$_i$ ($i=1,2$) represents a polarizing beam splitter (PBS) which transmits the $R$-polarized photon and reflects the $L$-polarized photon, respectively. HWP represents a half-wave
plate oriented at 0$^\circ$ and it is used to complete the unitary
transformation $\sigma_z=|R\rangle\langle R|-|L\rangle\langle L|$ on
a photon.
} \label{CNOT}
\end{center}
\end{figure}

Suppose  that the input state of the composite system composed of the flying single photon and the diamond NV center is
\begin{eqnarray}                    \label{eq7}
|\psi_{\text{in}}\rangle=(\cos\alpha|R\rangle_c+\sin\alpha|L\rangle_c)\otimes(\cos\beta|+\rangle_t+\sin\beta|-\rangle_t).
\end{eqnarray}
Here subscripts $c$ and $t$ represent the control qubit (the flying single photon) and the target qubit (the diamond NV center), respectively. First, the input single photon is split into two wave-packets by a polarizing beam splitter (PBS), say PBS$_1$. Second, the $R$-polarized component does not interact with the diamond NV center, whereas the $L$-polarized component interacts with the diamond NV center and then arrives at PBS$_2$ simultaneously with the $R$-polarized component. Third, before and after the photon interacts with the diamond NV center, a Hadamard operation $H_e$  is performed on the diamond NV center, respectively. Here $H_e$ completes the following transformations
\begin{eqnarray}                    \label{eq8}
|+\rangle
\xrightarrow{H_e}|\rightarrow\rangle\equiv\frac{1}{\sqrt{2}}(|+\rangle+|-\rangle),
\;\;\;\;\;\;\;\;\;\;\;\;
|-\rangle\xrightarrow{H_e}|\leftarrow\rangle\equiv\frac{1}{\sqrt{2}}(|+\rangle-|-\rangle).
\end{eqnarray}
Finally, a  single-qubit operation $\sigma_z=|R\rangle\langle R|-|L\rangle\langle L|$ is performed on the output  photon with a half-wave plate HWP oriented at $0^\circ$. With these operations, the state of the composite system evolves as follows:
\begin{eqnarray}                    \label{eq9}
\begin{split}
&|\psi_{\text{in}}\rangle\xrightarrow{H_e,\;\text{PBS}_1}(\cos\alpha|R\rangle_c+\sin\alpha|L\rangle_c)(\cos\beta|\rightarrow\rangle_{t}+\sin\beta|\leftarrow\rangle_{t})
\\&
\;\;\;\;\;\;\;\;\;\;\xrightarrow{\text{NV}}\cos\alpha|R\rangle_c(\cos\beta|\rightarrow\rangle_{t}+\sin\beta|\leftarrow\rangle_{t})
-\sin\alpha|L\rangle_c(\cos\beta|\leftarrow\rangle_{t}+\sin\beta|\rightarrow\rangle_{t})
\\&
\;\;\;\;\;\;\;\;\;\;\xrightarrow{H_e,\;\text{PBS}_2}\cos\alpha|R\rangle_c(\cos\beta|+\rangle_{t}+\sin\beta|-\rangle_{t})
-\sin\alpha|L\rangle_c(\cos\beta|-\rangle_{t}+\sin\beta|+\rangle_{t})
\\&
\;\;\;\;\;\;\;\;\;\;\xrightarrow{\text{HWP}}|\psi_{\text{out}}\rangle=\cos\alpha|R\rangle_c(\cos\beta|+\rangle_{t}+\sin\beta|-\rangle_{t})+\sin\alpha|L\rangle_c(\cos\beta|-\rangle_{t}+\sin\beta|+\rangle_{t}).
\end{split}
\end{eqnarray}
The quantum circuit shown in Fig. 2 completes the transformation $|\psi_{\text{in}}\rangle\xrightarrow{\text{CNOT}}|\psi_{\text{out}}\rangle$. That is, it implements a CNOT gate on a hybrid photon-NV system. If the flying  single photon is in state $|L\rangle$, the spins of the electron associated with the diamond NV center are flipped; otherwise, the spins of the electron remain unchanged.


{\bf Compact quantum circuit for implementing a Toffoli gate on a hybrid system.}
The principle of our hybrid Toffoli gate is shown in Fig. 3. This gate performs a CNOT operation on the two diamond NV centers, $\text{NV}_{c_2}$ and $\text{NV}_{t}$, when the flying single photon $c_1$ is in  state $|L\rangle$. Suppose that the system composed of $c_1$, NV$_{c_2}$, and NV$_{t}$ is prepared in the state
\begin{eqnarray}                    \label{eq10}
|\Phi_{\text{in}}\rangle=(\cos\alpha|R\rangle_{c_1}+\sin\alpha|L\rangle_{c_1})\otimes(\cos\beta|+\rangle_{c_2}+\sin\beta|-\rangle_{c_2})\otimes(\cos\delta|+\rangle_{t}+\sin\delta|-\rangle_{t}).\label{Toffoliin}
\end{eqnarray}
Our hybrid Toffoli gate works with the following steps.

\begin{figure}[h!]      
\begin{center}
\includegraphics[width=11.50 cm,angle=0]{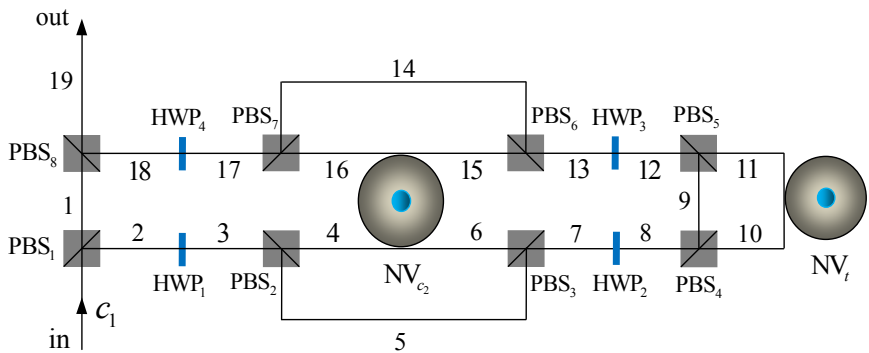}
\caption{ Compact quantum circuit for implementing a Toffoli gate which performs a CNOT operation on the two diamond NV centers if the flying single photon is in state $\vert L\rangle$.
}\label{Toffoli1}
\end{center}
\end{figure}

First, the $R$-polarized component of the input single photon $c_1$ is transmitted to  spatial mode 1 by PBS$_1$ and then arrives at PBS$_8$ directly, whereas the $L$-polarized component is reflected to  spatial mode 2  for interacting with the diamond NV centers. When the photon emits from spatial mode 2, it passes through the block composed of PBS$_2$, NV$_{c_2}$, and PBS$_3$, and a Hadamard operation $H_p$ is performed on it with a half-wave plate (HWP) oriented at 22.5$^\circ$ before and after it passes through the block, respectively.  We can obtain the following transformation induced by the above operations ($\text{PBS}_1\rightarrow\text{HWP}_1\rightarrow \text{PBS}_2\rightarrow\text{NV}_{c_2}\rightarrow \text{PBS}_3\rightarrow\text{HWP}_2$)
\begin{eqnarray}                  \label{eq11}
\begin{split}
|\Phi_{\text{in}}\rangle\rightarrow|\Phi\rangle_{1}=&\big[\cos\alpha|R_1\rangle_{c_1}(\cos\beta|+\rangle_{c_2}+\sin\beta|-\rangle_{c_2})
+\sin\alpha(\cos\beta|L_8\rangle_{c_1}|+\rangle_{c_2}\\&-\sin\beta|R_8\rangle_{c_1}|-\rangle_{c_2})\big](\cos\delta|+\rangle_{t}+\sin\delta|-\rangle_{t}).
\end{split}
\end{eqnarray}
Here and below, we use  $|R_i\rangle$ ($|L_i\rangle$) denotes the $R$- ($L$-) polarized  photon emitted from spatial mode $i$ ($i=1,2,\cdots, 19$).

Second, the photon passes through the block composed of PBS$_4$, NV$_t$, and PBS$_5$, and before and after the photon interacts with NV$_{t}$, an $H_e$ is performed on NV$_{t}$, respectively. These operations
($H_e\rightarrow\text{PBS}_4\rightarrow\text{NV}_t\rightarrow\text{PBS}_5\rightarrow H_e$) transform  $|\Phi\rangle_{1}$ into $|\Phi\rangle_{2}$. Here
\begin{eqnarray}                    \label{eq12}
\begin{split}
|\Phi\rangle_{2} =&
\big[\cos\alpha|R_{1}\rangle_{c_1}(\cos\beta|+\rangle_{c_2}+\sin\beta|-\rangle_{c_2})
+\sin\alpha\cos\beta|L_{12}\rangle_{c_1}|+\rangle_{c_2}\big]\\&\times(\cos\delta|+\rangle_t+\sin\delta|-\rangle_t)
-\sin\alpha\sin\beta|R_{12}\rangle_{c_1}|-\rangle_{c_2}(\cos\delta|-\rangle_t+\sin\delta|+\rangle_t).
\end{split}
\end{eqnarray}


Third, the photon emitting from spatial mode 12 passes through the block composed of PBS$_6$, NV$_{c_2}$, and PBS$_7$ . Before and after the photon passes through the block, an $H_p$ is  performed on it  with HWP$_3$  and HWP$_4$, respectively. After the wave-packet emitting from  spatial mode 18 arrives at PBS$_8$ simultaneously with the wave-packet emitting from spatial mode 1, the state of the system becomes
\begin{eqnarray}                    \label{eq13}
\begin{split}
|\Phi_{\text{out}}\rangle=&\big[\cos\alpha|R_{19}\rangle_{c_1}(\cos\beta|+\rangle_{c_2}+\sin\beta|-\rangle_{c_2})
+\sin\alpha\cos\beta|L_{19}\rangle_{c_1}|+\rangle_{c_2}\big]\\&\times(\cos\delta|+\rangle_{t}+\sin\delta|-\rangle_{t})
+\sin\alpha\sin\beta|L_{19}\rangle_{c_1}|-\rangle_{c_2}(\cos\delta|-\rangle_{t}+\sin\delta|+\rangle_{t}).\label{Toffoliout}
\end{split}
\end{eqnarray}

From Eqs. (\ref{Toffoliin})--(\ref{Toffoliout}), one can see that the quantum circuit in Fig. 3 completes the transformation $|\Phi_{\text{in}}\rangle\xrightarrow{\text{controlled-CNOT}}|\Phi_{\text{out}}\rangle$.
That is, it implements a Toffoli gate (it is also named a controlled-CNOT gate) which performs a CNOT operation on the two diamond NV centers when the control photon is in state $|L\rangle$; otherwise, the states of the two NV centers keep unchanged.


{\bf Quantum circuit for implementing a deterministic Fredkin gate on a hybrid system.}
Our Fredkin gate is used to exchange  the states of the two target diamond-NV-center-spin  qubits,  $\text{NV}_{t_1}$  and $\text{NV}_{t_2}$, when the flying single photon $c$ is in state $|L\rangle$; otherwise, the states of the two target qubits remain unchanged. The quantum circuit for implementing our Fredkin gate is shown in Fig. 4 and its  principle can be explained as follows.

\begin{figure}[!h]      
\begin{center}
\includegraphics[width=11 cm,angle=0]{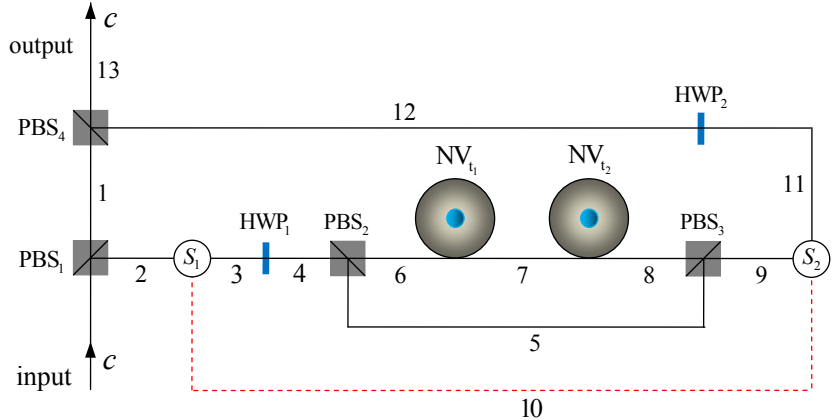}
\caption{ Quantum circuit for implementing a hybrid  Fredkin gate with a flying single photon as the control qubit and the two diamond NV centers as the target qubits.
} \label{Fredkin}
\end{center}
\end{figure}

Let us consider an input state of the three-qubit hybrid system composed of the control photon $c$ and the two target diamond NV centers $\text{NV}_{t_1}$  and $\text{NV}_{t_2}$,
\begin{eqnarray}                    \label{eq14}
\begin{split}
|\Xi_{\text{in}}\rangle=(\cos\alpha|R\rangle_{c}+\sin\alpha|L\rangle_{c})\otimes(\cos\beta|+\rangle_{t_1}+\sin\beta|-\rangle_{t_1})\otimes(\cos\delta|+\rangle_{t_2}+\sin\delta|-\rangle_{t_2}).
\end{split}
\end{eqnarray}
When the injecting control photon $c$ arrives at PBS$_1$, the state of the hybrid system is transformed from $|\Xi_{\text{in}}\rangle$ to $|\Xi_1\rangle$. Here
\begin{eqnarray}                    \label{eq15}
\begin{split}
|\Xi_1\rangle
=&(\cos\alpha|R_1\rangle_{c}+\sin\alpha|L_2\rangle_{c})(\cos\beta|+\rangle_{t_1}+\sin\beta|-\rangle_{t_1})(\cos\delta|+\rangle_{t_2}+\sin\delta|-\rangle_{t_2}).
\end{split}
\end{eqnarray}
The wave-packet emitting from  spatial mode 1  arrives at PBS$_4$ directly and the optical switch $S_1$ leads the wave-packet emitting from  spatial mode 2 to  spatial mode 3. After an $H_p$ is performed on the photon with HWP$_1$, it first passes through the block composed of PBS$_2$, NV$_{t_1}$, NV$_{t_2}$, and PBS$_3$, and then  arrives at $S_2$. $S_2$ leads the photon to spatial mode 10, followed with $S_1$ which leads the photon to spatial mode 3 for passing through HWP$_1$. These operations ($S_1\rightarrow\text{HWP}_1\rightarrow\text{PBS}_2\rightarrow\text{NV}_{t_1}\rightarrow\text{NV}_{t_2}\rightarrow\text{PBS}_3\rightarrow S_2\rightarrow S_1\rightarrow\text{HWP}_1$) transform the state of the hybrid system into
\begin{eqnarray}                    \label{eq16}
\begin{split}
|\Xi_2\rangle
=&\cos\alpha|R_1\rangle_{c}(\cos\beta|+\rangle_{t_1}+\sin\beta|-\rangle_{t_1})(\cos\delta|+\rangle_{t_2}+\sin\delta|-\rangle_{t_2})\\&
+\sin\alpha(\cos\beta\cos\delta|L_4\rangle_{c}|+\rangle_{t_1}|+\rangle_{t_2}
-\cos\beta\sin\delta|R_4\rangle_{c}|+\rangle_{t_1}|-\rangle_{t_2}\\&
-\sin\beta\cos\delta|R_4\rangle_{c}|-\rangle_{t_1}|+\rangle_{t_2}
+\sin\beta\sin\delta|L_4\rangle_{c}|-\rangle_{t_1}|-\rangle_{t_2}).
\end{split}
\end{eqnarray}

Before and after the second round, an $H_e$ is performed on each of NV$_{t_1}$ and NV$_{t_2}$. These operations ($H_{e_2},H_{e_3}\rightarrow\text{PBS}_2\rightarrow\text{NV}_{t_1}\rightarrow\text{NV}_{t_2}\rightarrow\text{PBS}_3\rightarrow H_{e_2},H_{e_3}\rightarrow S_2\rightarrow S_1$) transform $|\Xi_2\rangle$ into
\begin{eqnarray}                    \label{eq17}
\begin{split}
|\Xi_3\rangle =&
\cos\alpha|R_1\rangle_{c}(\cos\beta|+\rangle_{t_1}+\sin\beta|-\rangle_{t_1})(\cos\delta|+\rangle_{t_2}+\sin\delta|-\rangle_{t_2})\\&
+\sin\alpha
(\cos\beta\cos\delta|L_3\rangle_{c}|+\rangle_{t_1}|+\rangle_{t_2}
-\cos\beta\sin\delta|R_3\rangle_{c}|-\rangle_{t_1}|+\rangle_{t_2}\\&
-\sin\beta\cos\delta|R_3\rangle_{c}|+\rangle_{t_1}|-\rangle_{t_2}
+\sin\beta\sin\delta|L_3\rangle_{c}|-\rangle_{t_1}|-\rangle_{t_2}).
\end{split}
\end{eqnarray}

Next, the photon  passes through HWP$_1$ and the block composed of PBS$_2$, NV$_{t_1}$, NV$_{t_2}$, and PBS$_2$ in succession, and then $S_2$ leads it to spatial mode 11, followed with an $H_p$  (i.e., let it passes through HWP$_2$). Finally,  the wave-packet emitting from spatial mode 12 arrives at PBS$_4$ simultaneously with the wave-packet emitting from spatial mode 1. That is,  these  operations ($\text{HWP}_{1}\rightarrow\text{PBS}_2\rightarrow\text{NV}_{t_1}\rightarrow\text{NV}_{t_2}\rightarrow\text{PBS}_3\rightarrow S_{2}\rightarrow \text{HWP}_2\rightarrow\text{PBS}_4$) transform $|\Xi_3\rangle$ into
\begin{eqnarray}                    \label{eq18}
\begin{split}
|\Xi_{\text{out}}\rangle =&
\cos\alpha|R_{13}\rangle_{c}(\cos\beta|+\rangle_{t_1}+\sin\beta|-\rangle_{t_1})(\cos\delta|+\rangle_{t_2}+\sin\delta|-\rangle_{t_2})\\&
+\sin\alpha|L_{13}\rangle_{c}(\cos\beta|+\rangle_{t_2}+\sin\beta|-\rangle_{t_2})(\cos\delta|+\rangle_{t_1}+\sin\delta|-\rangle_{t_1}).
\end{split}
\end{eqnarray}

Putting all the pieces together, one can see that the quantum circuit shown in Fig. 4 completes the transformation $|\Xi_{\text{in}}\rangle\xrightarrow{\text{Fredkin}}|\Xi_{\text{out}}\rangle$. That is, the quantum circuit  shown in Fig. 4 implements a Fredkin gate which exchanges the spins of the two electrons associated with the diamond NV centers NV$_{t_1}$ and NV$_{t_2}$ when the flying single photon is in state $|L\rangle$; otherwise, the states of the two target qubits  remain  unchanged.


\section{Discussion}\label{sec3}

By far, several groups have experimentally demonstrated the coupling between a diamond NV center and a microcavity, such as microspheres\cite{microsphere1,microsphere2,microsphere3,microsphere4}, microdisks\cite{microdisk}, photonic crystals\cite{crystal1,crystal2,crystal3}, microtoroidal resonators\cite{toridal1,toridal2}, and fiber-based microcavity\cite{fiber-based}. It is a challenge to achieve the strong coupling between the NV and the cavity in experiments with current technology. Fortunately, the strong coupling between  NV centers in diamond nanocrystals and a whispering gallery mode (WGM) in a silica microsphere has been achieved\cite{microsphere1}. Larsson \emph{et al.}\cite{microsphere2} showed that it is possible to achieve the strong coupling between  NV centers in a diamond nanopillar coupled to a WGM in a silica microsphere. In 2013, Teissier \emph{et al.}\cite{mechanical} realized  an exceeding 10 MHz coupling strength between an  NV center and a diamond mechanical oscillator. In 2006, Park \emph{et al.}\cite{microsphere1} observed the strong coupling ($\emph{\text{g}}/2\pi=55$ MHz, $\gamma/2\pi=25$ MHz, $\kappa/2\pi=50$ MHz) in a diamond NV center coupled to a WGM in a silica microsphere.  Barclay \emph{et al.}\cite{Barclay} showed  that  the strong coupling with the parameters $[\emph{\text{g}},\kappa,\gamma_{\rm{tot}}]/2\pi=[2.25,0.16,0.013]$ GHz is possible in an NV nanocavity. In 2009, Barclay \emph{et al.}\cite{microdisk} showed that the parameters $[\emph{\text{g}},\kappa,\gamma,\gamma_{\rm{ZPL}}]/2\pi=[0.30,26,0.013,0.0004]$ GHz can be achieved in experiment for coupling the NV centers in single crystal diamond to an chip-based microcavity. Here $\gamma_{\rm{ZPL}}$ is the spontaneous emission rate of a diamond NV center into the zero phonon line (ZPL). For NV-microtoridal resonators, $|r(\omega_p)|\sim1$ can be achieved when $\emph{\text{g}}=2\pi\times500$ MHz with $\kappa=2\pi\times10$ GHz or $\kappa=2\pi\times1$ GHz\cite{NV-NV3}.

Our schemes work for the degenerate cavity modes, and it can be achieved by employing microtoroidal resonators\cite{toridal1,toridal2,degenerate2,Lei}, H1 photonic crystals\cite{unpolarized-photon1,unpolarized-photon2}, micropillars\cite{unpolarized-pillar1,unpolarized-pillar2,unpolarized-pillar3}, or fiber-based\cite{fiber-based} cavities. Our schemes are deterministic in principle. Our schemes have  high fidelities and efficiencies if the photon loss caused by the linear optics are not taken into account. Certainly, we should take the photon loss into account in the practical applications\cite{Bull1} as there are  the cavity absorption and scattering, and the absorption  from linear optical elements (such as the fibers, PBS, and HWP). Different  to  the protocol for generating entanglement between two NV centers\cite{threemeters}, our gates cannot be heralded by the destructive detection of a single photon. Our schemes can be inferred by the successful instances in postselection in practical applications of our gates.
For example, when our hybrid gate is used for quantum information transfer, the successful transfer of the information from the NV electron spin to the single photon polarization indicates the success of our CNOT gate. In principle, the photon loss can be reduced by improving experiment techniques and fabrication processing. The ZPL emission of an NV center is only 3\%-4\% of the total emission. In 2011, Barclay \emph{et al.}\cite{PRX} enhanced the ZPL emission of an NV center in a WGM nanocavity from $\sim$3\% to $\sim$16\%. Subsequently, they\cite{toridal1} enhanced the ZPL emission of  an NV center coupled to a microresonator from 3/100 to 36/133. In 2012, Faraon \emph{et al.}\cite{Faraon} enhanced the ZPL emission by a factor of $\sim$70  in photonic crystal cavities.

Fluctuations in the frequency of the optical transition of NV centers, due to the fluctuation in the charge environment, is a hurdle for our schemes. This spectral diffusion in the nanocavity devices results in an overall line width  which  can be much larger than the NV transition line width (13-16 MHz). Therefore, as that done by  Delft's group\cite{threemeters}, we should first  check the transition frequency of the NV centers before our schemes.
Spectral diffusion can be reduced by active stabilization technique, preselection of the transition frequency technique, or combination of high temperature annealing and subsequent surface treatment technique\cite{Diffusion1,Diffusion2,Diffusion3}. The optical transition frequencies of the two NV centers in our schemes for Toffoli and Fredkin gates can be tuned into resonance with each other by  applying an external electric field\cite{electricfield}.

 Our schemes work not only for the two-fold sublevels encoded for the electron-spin qubits but also for the non-degenerate spin sublevels lifted by a small external magnetic field. Dr\'{e}au \emph{et al.}\cite{anticrossing} demonstrated that the excited states occur sublevels anticrossing when $B\approx510$ G and the one for the ground states when $B\approx1020$ G. The state $A_2$ is robust against a relatively small magnetic field. For the non-degenerate one, if only the $R$-polarized photon matches the resonance transition, our schemes can implement the CNOT, Toffoli, and Fredkin gates only with a little modification on the quantum circuit in Fig. 2.

Compared with the parity-measurement approach in \cite{PCG1,PCG2} and the one  based on control path and merging gates\cite{Kerr1}, the auxiliary qubits are not required in our schemes, and the number of the nonlinear interactions required for our CNOT gate is fewer than that in \cite{PCG1,PCG2,Kerr1}. The complexity of our Toffoli and Fredkin gates beat their synthesis procedures in terms of two-qubit entangling gates largely as the well known cost of the Toffoli and Fredkin gates\cite{Toffolicost,Kerr1,Fredkincost} are six CNOT gates and five two-qubit entangling gates, respectively.

In summary, we have presented compact quantum circuits for the hybrid universal quantum gates assisted by the input-output process of a single photon. Our CNOT, Toffoli, and Fredkin gates work with the single-photon polarizations as the control qubits and the electron spins associated with the diamond NV centers as the target qubits.  Our schemes take the advantages of the theoretical and experimental progress in the fast electron-spin manipulation, the long-lived electron-spin coherence time, and the flexible controllability of the single photon. All our schemes are compact, economic, and simple. They have high fidelities and efficiencies  with current technology.


\section{Methods}\label{sec4}

{\bf Average fidelities and efficiencies of the gates.}
We use the fidelity and the efficiency to characterize the performance of our universal quantum gates. In order to characterize the construction of these gates, we specify the evolutions of the hybrid systems from the initial sates $|\psi_{\text{in}}\rangle$  to the output states $|\psi_{\text{out}}\rangle$ in the ideal case. The fidelity of a quantum gate is defined as $F=|\langle\psi_{\text{out}}|\psi_{\text{out}}'\rangle|^2$, and it
is the probability that the normalized output state of the whole system in the ideal case $|\psi_{\text{out}}\rangle$ overlaps with the realistic state $|\psi_{\text{out}}'\rangle$. Taking the CNOT gate as an example, in the ideal case (i.e., $r(\omega_p)\simeq1$ and $r_0(\omega_p) =-1$), the normalized output state of our scheme is given by Eq. (\ref{eq9}), that is,
\begin{eqnarray}                    \label{eq19}
|\psi_{\text{out}}\rangle_{CT}\;=\;\cos\alpha|R\rangle_{c}(\cos\beta|+\rangle_{t}+\sin\beta|-\rangle_{t})
+\sin\alpha|L\rangle_{c}(\cos\beta|-\rangle_{t}+\sin\beta|+\rangle_{t}).
\end{eqnarray}
By substituting Eq. (\ref{eq5}) for Eq. (\ref{eq6}) and combing the evolutions of the  state for the CNOT gate,  the non-normalized output state in the realistic case becomes
\begin{eqnarray}                    \label{eq20}
\begin{split}
|\psi_{\text{out}}'\rangle_{CT} &=
\cos\alpha|R\rangle_{c}(\cos\beta|+\rangle_{t}+\sin\beta|-\rangle_{t})
 +\frac{\sin\alpha}{2}|L\rangle_{c}\big[\cos\beta(|r|+1)-\sin\beta(|r|-1)\big]|-\rangle_{t}\\&\;\;\;\;\;
  +\frac{\sin\alpha}{2}|L\rangle_{c}\big[\sin\beta(|r|+1)-\cos\beta(|r|-1)\big]|+\rangle_{t}.
\end{split}
\end{eqnarray}
That is, the average fidelity of our CNOT gate can be expressed as
\begin{eqnarray}                    \label{eq21}
\overline{F}_{CT}=\frac{1}{4\pi^2}\int_0^{2\pi}d\alpha\int_0^{2\pi}d\beta|\langle\psi_{\text{out}}|\psi_{\text{out}}'\rangle|^2.
\end{eqnarray}
Using the same arguments for the CNOT gate, one can obtain the average fidelities of the Toffoli gate $\overline{F}_T$ and  the Fredkin gate $\overline{F}_F$, shown in Fig. 5a.

\begin{figure}[!h]           
\begin{center}
\includegraphics[width=6.2 cm,angle=0]{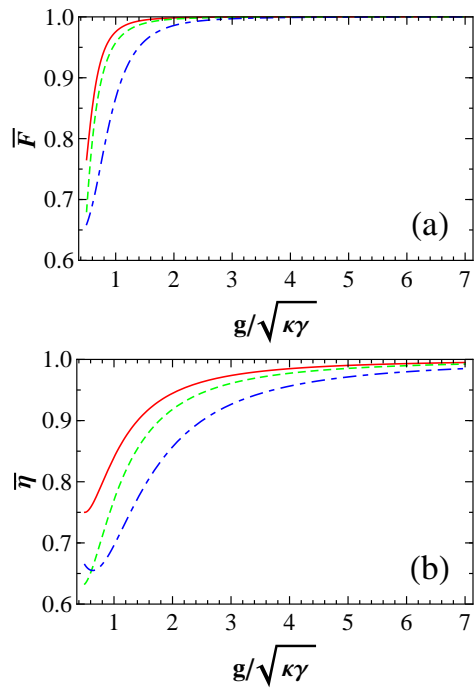}
\caption{The average fidelities ($\overline{F}$) and the average efficiencies ($\overline{\eta}$) of our universal quantum gates on photon-NV hybrid systems vs $\emph{\text{g}}/\sqrt{\kappa\gamma}$. Here the red solid line, the green dashed line, and the blue dash-dotted line correspond to those of our CNOT, Toffoli, and Fredkin gates, respectively. $\emph{\text{g}}/\sqrt{\kappa\gamma}\geq 0.5$.
} \label{Fidelity}
\end{center}
\end{figure}

Since the flying single photon may be lost during the operation for a gate, we can use $\eta=n_{\text{output}}/n_{\text{input}}$ to characterize the efficiency of a gate. Here $n_{\text{input}}$ and $n_{\text{output}}$ are the numbers of the input photons and the output photons, respectively. Combing the spin-selection rules in the realistic case described by Eq. (\ref{eq5}) and the evolutions of the system from the input states to the output states, the average efficiencies of our gates, averaged over  $\alpha,\beta,\delta\in[0,2\pi]$, can be obtained as follows:
\begin{eqnarray}                    \label{eq22}
\overline{\eta}_{CT}=\frac{3 + |r|^2}{4},
\end{eqnarray}
\begin{eqnarray}                    \label{eq23}
\overline{\eta}_{T}=\frac{(3 + |r|^2) (27 + 2 |r| + 4 |r| ^2 - 2 |r| ^3 + |r|^4)}{128},
\end{eqnarray}
%
%
\begin{eqnarray}                    \label{eq24}
\begin{split}
\overline{\eta}_{F}=&\frac{1361 - 156 |r| + 286 |r| ^2 + 28 |r| ^3 + 239 |r| ^4 + 152 |r| ^5+148 |r| ^6 -
 24| r| ^7 - | r| ^8 + 4 |r| ^9 + 14 |r| ^{10} - 4 |r| ^{11} + |r| ^{12}}{2048}.
\end{split}
\end{eqnarray}
The average efficiencies of our gates vary  with $\emph{\text{g}}/\sqrt{\kappa\gamma}$, shown in Fig. 5b.

{\bf The feasibility of the gates.}
The fidelities of our gates can be reduced by the few percent by the experimental operation imperfection, such as electronic spin preparation with a low limit fidelity of $99.7\pm 0.1\%$ to   $m_s=0$  and   $99.2\pm 0.1\%$ to $m_s=\pm 1$\cite{readout3}. Bernien \emph{et al.}\cite{threemeters} showed that the fidelity of their setup can be reduced by the microwave pulse errors ($\sim$3.5\%), off-resonant excitation errors ($\sim$1\%), spin
decoherence ($<$1\%), the charge fluctuation due to the optical frequencies, and spin-flip errors in the excited states during the optical excitation ($\sim$1\%). Togan \emph{et al.}\cite{photon-NV} pointed out that the fidelity can be reduced by the imperfect optical transitions due to the moderate and high strain, the path length fluctuation ($\sim$4\%), and the signal to noise ratio in the ZPL channel ($\sim$11\%). The charge fluctuation and the imperfect electron-spin population can be decreased by exploiting a repeated-until-success  (the negative charge state and on resonance) fashion\cite{threemeters} before performing our gates.

\section*{Acknowledgments}

 This work was supported by the National Natural Science Foundation of China under Grant
Nos. 11175094 and 91221205,  the National Basic Research Program
of China under Grants No. 2009CB929402 and No. 2011CB9216002, and the China Postdoctoral Science Foundation under
Grant No. 2014M550703. GLL is
a member of the Center of Atomic and Molecular Nanosciences,
Tsinghua University.

\end{document}